\documentclass[english,aps,prx,superscriptaddress,preprint]{revtex4}
\usepackage[T1]{fontenc}
\usepackage[latin9]{inputenc}
\usepackage{amsmath}
\usepackage{graphicx}
\usepackage{amssymb}
\usepackage{esint}

\makeatletter
\@ifundefined{textcolor}{}
{%
 \definecolor{BLACK}{gray}{0}
 \definecolor{WHITE}{gray}{1}
 \definecolor{RED}{rgb}{1,0,0}
 \definecolor{GREEN}{rgb}{0,1,0}
 \definecolor{BLUE}{rgb}{0,0,1}
 \definecolor{CYAN}{cmyk}{1,0,0,0}
 \definecolor{MAGENTA}{cmyk}{0,1,0,0}
 \definecolor{YELLOW}{cmyk}{0,0,1,0}
 }

\usepackage{dcolumn}\usepackage{bm}

\makeatother

\usepackage{babel}

\begin{document}

\title{On the quantitative calculation of the cosmological constant of the quantum vacuum}

\author{Hongwei Xiong}
\email{xionghw@zjut.edu.cn}

\address{College of Science, Zhejiang University of Technology, Hangzhou 310023, China}

\date{\today}

\begin{abstract}

It is widely believed that as one of the candidates for dark energy, the cosmological constant should relate directly with the quantum vacuum. Despite decades of theoretical effects, however, there is still no quantitative interpretation of the observed cosmological constant. 
In this work, we consider the quantum state of the whole universe including the quantum vacuum. Everett's relative-state formulation, vacuum quantum fluctuations and the validity of Einstein's field equation at macroscopic scales imply that our universe wave function might be a superposition of states with different cosmological constants. In the density matrix formulation of this quantum universe, the quasi-thermal equilibrium state is described by a specific cosmological constant with the maximum probability. Without any fitting parameter, the ratio between the vacuum energy density due to the cosmological constant (dark energy) and the critical density of the universe is $68.85\%$ based on simple equations in our theoretic model, which agrees very well with the best current astronomical observations of $68.5\%$.


Subject Areas: Cosmology, Gravitation, Quantum Physics

\end{abstract}

\maketitle

\section{Introduction}

After two decades of study, the answer to the physical mechanism of the accelerating universe is as mysterious as ever \cite{SA}. Dark energy is the dominating hypothesis since the 1990s to explain the observations of the expanding universe at an accelerating rate \cite{acceleration,acceleration1}. The cosmological constant is one of the most promising candidate for dark energy \cite{weinberg,cos}. Other proposals include the scalar field model such as quintessence \cite{quin}. Even astronomical observations in future verify the cosmological constant model with much higher accuracy, it seems that to resolve the puzzle of dark energy, we need some new physical pictures that will significantly change our cosmic view.

Including the cosmological constant term, the Einstein's field equation is written as
 \begin{equation}
 G_{\mu\nu}=8\pi G(T_{\mu\nu}+T_{\mu\nu}^{\Lambda}).
 \label{Einstein}
 \end{equation}
We will use the units with $c=1$ throughout. The left side is the Einstein tensor representing the geometry of spacetime.
$T_{\mu\nu}$ is the stress-energy tensor of the material contents of the universe including particles, dark matter and radiation. The stress-energy tensor of the quantum vacuum is
\begin{equation}
T^{\Lambda}_{\mu\nu}=-\rho_{\Lambda}g_{\mu\nu},
\end{equation}
where $\rho_\Lambda$ represents the vacuum energy density. $\rho_\Lambda=\Lambda/8\pi G$ is proportional to the Einstein's cosmological constant $\Lambda$. The conservation of the stress-energy tensor of the quantum vacuum ($D^{\mu}T_{\mu\nu}^{\Lambda}=0$) agrees with the property of the cosmological constant that $\Lambda$ is uniform and time-independent. The vacuum energy density in a local inertial frame is $\Lambda/8\pi G$, which is invariant under the Lorentz transformation for this local inertial frame. 

$T_{\mu\nu}^{\Lambda}$ may have both geometrical interpretation and energy-momentum interpretation. At present, the mainstream opinion is that the cosmological constant term originates from quantum fluctuations in the vacuum, firstly addressed by Zel'dovich \cite{Zeldovich,Zeldovich1}. Despite many years of studies, we are still bewildered by the fine tuning problem and coincidence problem, i.e., the dark energy density is not only at odds with all possible fundamental energy scales and requires therefore fine tuning, but also that this particular value is almost identical to a seemingly unrelated number, the present matter energy density.

The purpose of the present work is to calculate the cosmological constant in a quantitative way. We will study the cosmological constant by the assumption that we should extend the wave function of the universe to include the quantum vacuum. We will propose two postulates based on the general discussion on the evolution of the quantum universe, and then calculate the cosmological constant based on these two postulates. The paper is organized as follows. In Sec. II, we give the first postulate based on the consideration of the spontaneous symmetry breaking of the quantum vacuum. In Sec. III, we give the density matrix and qualitative picture of the universe. In Sec. IV, we give the second postulate to consider the quasi-thermal equilibrium condition between the quantum vacuum and the material contents of the universe. In this section, we also provide the theoretical model to calculate the effective energy for matter, radiation and vacuum which will be used in Sec. V to calculate numerically the cosmological constant with the maximum probability. In the last section, we give a summary of our work and a further discussion on the whole cosmic history, beginning from a pure quantum vacuum without matter and radiation.

\section{The postulate about quantum vacuum}

In addition to the fine tuning and coincidence problems mentioned in previous section, there is another conundrum due to vacuum quantum fluctuations. 
It is well known that the quantum world is never still. When quantum gravity is considered, the curvature of spacetime and even its structure would be subject to fluctuations. As envisioned by Wheeler in 1957 \cite{Wheeler}, the quantum vacuum becomes increasing chaotic as smaller regions of space are considered. At the scale of the Planck length $l_P$, even the topology of space would undergo violent fluctuations. Assume that the energy fluctuation $\Delta E_P$ contained within a Planck volume $l_P^3$ is of the order of the Planck energy $E_P$. If the interaction and correlation between neighboring quantum vacua are omitted, the overall vacuum energy fluctuations within a region of volume $\Delta V$ would be $\Delta E\sim \sqrt{\Delta V/l_P^3}E_P$. However, this will immediately lead to a great difficulty. We consider an atom with radius of about 1\AA~as an example. Simple calculations show that within the atomic size, $\Delta E$ is more than 55 orders of magnitude larger than the rest mass of a proton! This leads to an unavoidable problem that why the spacetime is usually smooth at atomic and macroscopic scales. 

It is clear that there should be interaction and correlation between neighboring quantum vacua. Although we do not know the details, it is not unreasonable to imagine that the correlated interaction will lead to a spontaneous symmetry breaking of the quantum vacuum, similarly to the crystallization of atomic or molecular gases. For atomic or molecular gases in  a closed container, before the formation of a crystal slab, the particle number fluctuations within a region of volume $\Delta V$ are $\Delta N\sim \sqrt{\overline{n}\Delta V}$ with $\overline n$ being the average particle number density. After the formation of the crystal slab, it seems that the particle number fluctuations within the crystal slab could be negligible because of the formation of stable solid forms. 
Rigorously speaking, however, there should be particle number fluctuations.
For this closed system that the environment coupling is negligible, after the formation of the crystal slab and before an observation, the many-body wave function is a superposition of the wave functions of the crystal slab with different shapes and at different locations. There would be still significant particle number fluctuations when all these crystal slabs in different parallel worlds are considered. We know that the spontaneous symmetry breaking plays a key role in the formation of the crystal slab and provides the clue to solve the problem of particle number fluctuations. The ferromagnetic phase transition below the Curie temperature provides another clear example of the spontaneous symmetry breaking that there would be macroscopic magnetic moment with random direction in different parallel worlds. However, there are still significant fluctuations of the magnetic moment if all these parallel worlds are considered, although in any parallel world the macroscopic magnetic moment has a definite direction. We know that the simplified Ising spin model provides a beautiful picture to understand the ferromagnetic phase transition by considering the interaction of adjacent spins.

Inspired by the above discussions, our first postulate of the quantum vacuum is proposed as follows.

{\bf Postulate 1}: 
When the interaction and correlation between fluctuating and neighboring quantum vacua are considered, the quantum vacuum is assumed as the superposition of stable vacuum states with different cosmological constants, which is written as
\begin{equation}
\left|\Psi_{vacuum}\right\rangle=\sum_\Lambda\alpha(\Lambda)\left|\Lambda\right\rangle+\beta(t)\left|\Psi_{residual}\right\rangle.
\end{equation}

Here $|\Lambda\rangle$ denotes the quantum vacuum state with the cosmological constant $\Lambda$.
The average value and fluctuations of $\Lambda$ are then $\overline\Lambda=\sum_\Lambda|\alpha(\Lambda)|^2\Lambda$ and $\delta \Lambda=\sqrt{\sum_\Lambda|\alpha(\Lambda)|^2(\Lambda-\overline\Lambda)^2}$, respectively. Although $\Lambda$ is a constant in a specific universe described by $\left|\Lambda\right\rangle$, there can be significant fluctuations of the vacuum energy. The second term $\beta(t)\left|\Psi_{residual}\right\rangle$ represents the wave function of the quantum vacuum not described by $\sum_\Lambda \alpha(\Lambda)|\Lambda \rangle$, because $\left|\Lambda\right\rangle$ can not be the exact eigenstate and $\{\left|\Lambda\right\rangle\}$ should not be the complete orthonormal eigenstates due to the complexity of the correlated interactions in the quantum vacuum. Nevertheless, we assume that $|\beta(t)|^2<<1$ at the present time of our universe. We will show in due course that  this term will play a key role in the evolution of our universe, which distinguishes our theory from other proposals about multiverse with different cosmological constants \cite{multiverse,Susskind}.

To show clearly the meaning of Postulate 1, we consider several applications of this postulate as follows.

As the first application of Postulate 1, we consider a thought experiment of the evolution of an initial quantum state which is the product state of the quantum vacuum and a nebula, i.e., $\left|\Psi(t=0)\right\rangle=\sum_\Lambda\alpha(\Lambda)\left|\Lambda\right\rangle\otimes\left|\Psi_{nebula}\right\rangle$. If the second term $\beta(t)\left|\Psi_{residual}\right\rangle$ of the quantum vacuum is omitted for the time being, and omitting the counteraction to the quantum vacuum by the nebula, the interaction between quantum vacuum and the nebula leads naturally to the following entangled wave function
\begin{equation}
\left|\Psi(t)\right\rangle=\sum_\Lambda\alpha(\Lambda)\left|\Lambda\right\rangle\otimes\left|\Psi_{nebula}(\Lambda, matter, t)\right\rangle.
\end{equation}
Here $\left|\Psi_{nebula}(\Lambda,matter,t)\right\rangle$ means that the evolution of the nebula is dependent on the value of $\Lambda$ in different parallel worlds. After billions of years, in the parallel world of appropriate cosmological constant, there would be formation of stable structure and even life to understand the universe. For a specific observer, what he/she observes is a specific cosmological constant based on the above entangled quantum state. Of course, the above initial product state is only a thought experiment. In reality, because of the big bang origin of our universe, the initial state is not a product state of the quantum vacuum and the material contents of our universe.

As a further application of Postulate 1, now we consider the universe quantum state experienced by an observer which is also a many-body quantum state. Generally speaking, the quantum state of the observer can be written as $\left|\Psi_o\right\rangle=\sum_\Lambda \gamma(\Lambda)\left|\Lambda,matter\right\rangle$. In addition, $\left|\Psi_{universe}\right\rangle$ is assumed as the whole universe wave function including the vacuum. Application of quantum theory to the universe is not a new idea. In the 1960s, partly inspired by Everett's work \cite{Everett} on the universe wave function, the foundation of quantum cosmology was built by DeWitt, Misner and Wheeler \cite{Dewitt,Misner}. Quantum mechanics has become an indispensable element to understand the origin of the universe, e.g., the work about the wave function of the universe by Hartle and Hawking \cite{Hawking}. We will show in the present work that quantum mechanics may play important role in the whole cosmic evolution.

Based on the relative state formulation \cite{Everett} which was used by Everett to solve the measurement problem in quantum mechanics with his many-worlds interpretation, the quantum state the observer experiences is
\begin{equation}
\Psi_{rel}^{o}=A\sum_i\langle\psi_i|\otimes\langle\Psi_o|\Psi_{universe}\rangle\psi_i.
\end{equation}
Here $A$ is a normalization constant, and $\{|\psi_i\rangle\}$ is a complete orthonormal basis of the universe besides the observer. It is not difficult to prove that the relative state function $\Psi_{rel}^{o}$ is unique \cite{Everett} once the wave function of the whole universe and the observer are known, i.e., it does not depend on the choice of $\{|\psi_i\rangle\}$.

It is clear that a stable observer (at least with our present understanding of life) should not be a quantum supposition of significantly different $\Lambda$, i.e., it is reasonable to assume the wave function of the observer as $\left|\Psi_o\right\rangle=\left|matter,\Lambda_o\right\rangle$ with $\Lambda_o$ being a constant in the region encircling the observer. From the above relative-state formulation and Postulate 1, we see that the whole universe the observer experiences has the same vacuum energy density determined by $\Lambda_o$. Hence, we see that Postulate 1 is in agreement with Einstein's field equation including the cosmological constant term. In fact, Postulate 1 is proposed based on four considerations: (1) The wave function is used to describe the whole universe including the vacuum and its quantum fluctuations; (2) Einstein's field equation including the cosmological constant term should be the macroscopic correspondence of the quantum description of the quantum vacuum; (3) Everett's relative state formulation to describe the world by a stable observer; (4) Spontaneous symmetry breaking due to the interaction between neighboring quantum vacua. With these considerations, it is almost logical to get Postulate 1. Of course, different observers may experience completely different cosmological constant and even find different cosmic evolution because Einstein's field equation can be applied to any universe with different cosmological constant.

\section{Density matrix and qualitative picture of the universe }

It is natural to consider the density matrix formulation of our universe based on Postulate 1. Assume the initial unknown quantum state is $\left|\Psi_{universe}(t=0)\right\rangle$. After some time (maybe after the end of the inflation of the universe), when the stable quantum vacuum is formed, based on Postulate 1, the quantum state of the universe can be written as
\begin{equation}
\left|\Psi_{universe}(t)\right\rangle=\sum_\Lambda\alpha(\Lambda,t)\left|\Psi_{vacuum}(\Lambda,t)\right\rangle\otimes\left|\Psi_{matter}(\Lambda,t)\right\rangle+\beta(t)\left|\Psi_{residual}\right\rangle.
\end{equation}

We should stress again that because of the complexity of the interaction between the material contents and quantum vacuum, and the interaction between neighboring quantum vacua, $\{\left|\Psi_{vacuum}(\Lambda,t)\right\rangle\otimes\left|\Psi_{matter}(\Lambda,t)\right\rangle\}$ can not be the complete orthonormal basis of the whole universe, although Postulate 1 strongly suggests that it may give the main contribution to $\left|\Psi_{universe}(t)\right\rangle$. The residual contribution is given by the second term in the above equation. $|\beta(t)|^2$ may be much smaller than 1, while the vacuum energy described by $\left|\Psi_{residual}\right\rangle$ can be both time dependent and location dependent which is determined by the initial unknown universe wave function and the whole Hamiltonian of our universe. It is of course impossible to calculate the evolution of $\beta(t)$ and $\left|\Psi_{residual}\right\rangle$. If $\{\left|\Psi_{vacuum}(\Lambda,t)\right\rangle\otimes\left|\Psi_{matter}(\Lambda,t)\right\rangle\}$ were the rigorous complete orthonormal basis of the cosmic Hamiltonian, the linear superposition principle of quantum mechanics means that $|\alpha(\Lambda,t)|^2$ should be time independent. However, based on the standard quantum mechanics, the existence of the term $\beta(t)\left|\Psi_{residual}\right\rangle$ implies that $|\alpha(\Lambda,t)|^2$ should be time dependent. In other words, the term $\beta(t)\left|\Psi_{residual}\right\rangle$ and the Hamiltonian of the whole universe will induce the transition between quantum states $\left|\Psi_{vacuum}(\Lambda,t)\right\rangle\otimes\left|\Psi_{matter}(\Lambda,t)\right\rangle$ with different $\Lambda$.

For the observable universe, the appropriate description is the reduced density matrix $\rho_{universe}$ obtained by the partial trace over the component we can not observe because of the finite velocity of information propagation. 
Under this consideration, we have
\begin{equation}
\rho_{universe}\simeq\sum_\Lambda|\alpha(\Lambda,t)|^2|\Psi_{vacuum}(\Lambda,t)\rangle\otimes|\Psi_{matter}(\Lambda,t)\rangle\langle\Psi_{vacuum}(\Lambda,t)|\otimes\langle\Psi_{matter}(\Lambda,t)|.
\label{densitymatrix}
\end{equation}

This density matrix formulation reminds us the statistical thermal equilibrium and gives us strong suggestion that we can get specific value of $\Lambda$ for a real observer based on the thermal equilibrium evolution of our universe. Here, the thermal equilibrium refers to the thermal equilibrium between the quantum vacuum and the material contents of our universe. Because of the expansion of our universe, rigorously speaking, this is a quasi-thermal equilibrium process. This quasi-thermal equilibrium expansion process of the universe provides another mechanism for the time dependency of $|\alpha(\Lambda,t)|^2$. The application of statistical physics tells us that 
the maximum value of $|\alpha(\Lambda,t)|^2$ will give the definite value of $\Lambda$ for the real observer. 

Omitting for the time being the radiation in the universe, when the coupling and quasi-thermal equilibrium between matter and quantum vacuum are considered, it is natural that the vacuum energy density is of the same order of the matter energy density, which gives us chance to solve the fine-detuning problem and coincidence problem. However, one may reject this proposal by pointing out that this means that the vacuum energy density would take the same time dependence as the matter energy density, which contradicts with our observations. A little thought based on Postulate 1 will show that there is no such contradiction because at different time of the universe, the observer will observe different cosmological constant which is proportional to the matter energy density at that time, determined by the quasi-thermal equilibrium condition at that time. When the observer collects the information of the evolution of the universe, based on the application of the relative state formulation, the observer will only find uniform and time independent vacuum energy density for the evolution history of the universe.  In Fig. 1, we show further this physical picture of the evolution of our universe and the cosmic history by a real observer.

\begin{figure}
 \centering \includegraphics[width=0.8\textwidth]{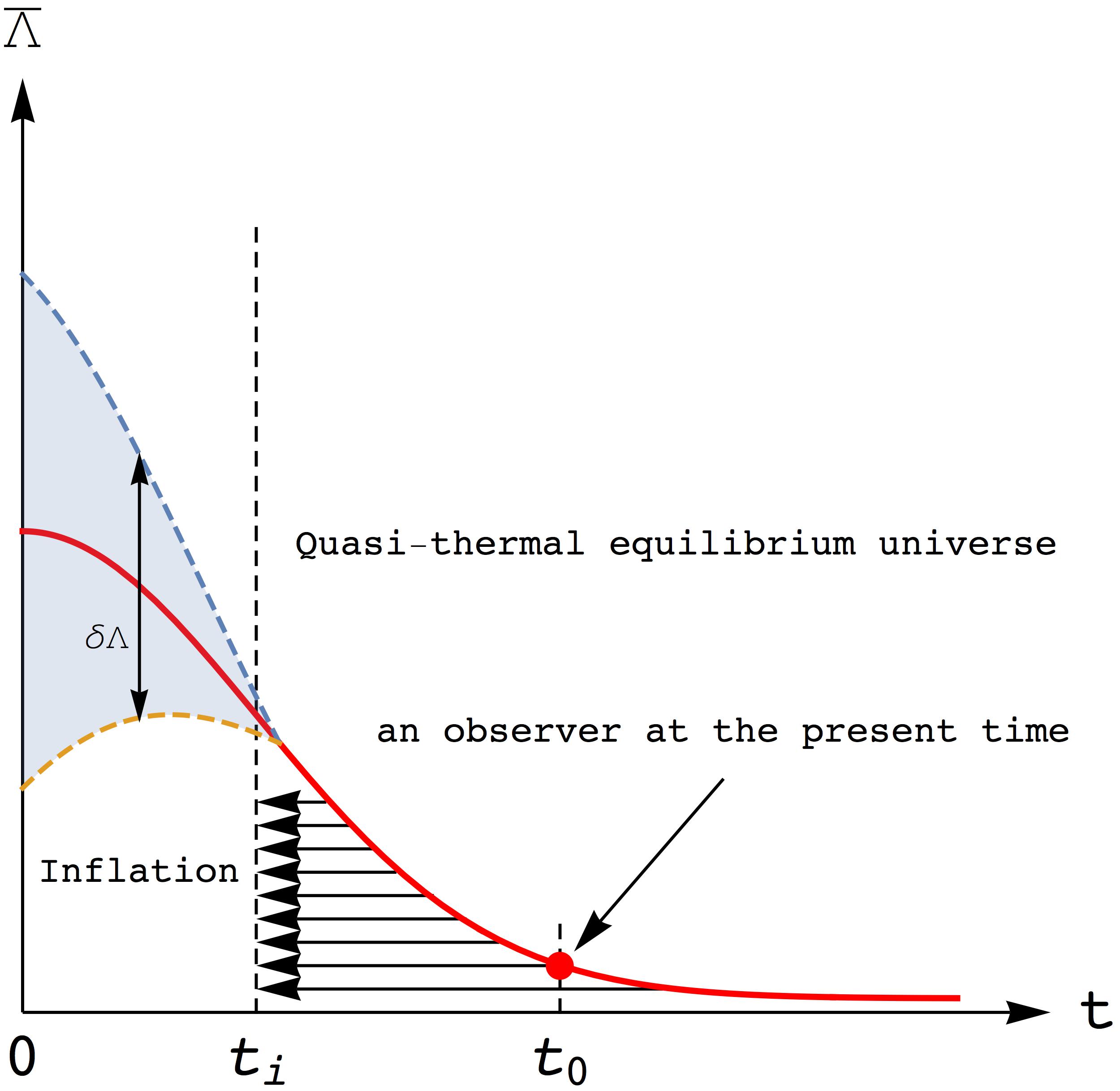} 
 \caption{A schematic diagram of the cosmic evolution based on our model. The red solid line denotes the average value of the cosmological constant as a function of time. During the inflationary stage, the fluctuation $\delta\Lambda$ of the cosmological constant shown by the blue region can be of the order of the average cosmological constant. In this inflationary stage, the vacuum energy leads to the exponential expansion of the universe, while the fierce vacuum fluctuations cause the creation of the material contents. The end of the inflation is due to the gradual establishment of the quasi-thermal equilibrium between quantum vacuum and material contents. In the quasi-thermal equilibrium stage of our universe, the real observer exists in the universe of the cosmological constant with the maximum probability. This cosmological constant with the maximum probability is time dependent. However, at the present time $t_0$, the relative-state formulation makes the cosmological constant become time-independent in the cosmic history observed by the observer, illustrated by the black solid line with an arrow. The red solid disk represents a real observer at the present time $t_0$. We stress that the evolution history of the average cosmological constant (red solid line) and that of a real observer (black solid line with arrow) is different.}
\end{figure}

\section{Quasi-thermal equilibrium condition between quantum vacuum and material contents of the universe}

To calculate quantitatively the vacuum energy density experienced by a real observer at different cosmic time, we first give the second postulate of the present work.

{\bf Postulate 2}: The quasi-thermal equilibrium between the material contents and the quantum vacuum is determined by the assumption that, in a local inertial frame comoving with the expansion of the universe, the effective overall energy experienced by a "detector" inside the spacetime for the vacuum energy equals to that of all the material contents in the universe.

To show clearly the meaning of this postulate, we will first calculate the effective overall energy for matter, radiation and vacuum energy as follows.

\textit{(i) The effective overall energy of matter} 

Firstly, we consider the matter situation in a Minkowski spacetime or a local inertial frame in curved spacetime. We consider a thought experiment that there is a sphere of radius $r$ covered with fictitious detectors on the spherical surface to measure the energy of a particle with rest mass $m$. At time $t=0$, the particle with velocity $\bf v$ arrives at the center of the sphere. Special relativity tells us that the relativistic energy of this particle is $m/\sqrt{1-v^2}$. In principle, the presence of the particle will induce vacuum excitations in the spacetime \cite{MyWork}, and these vacuum excitations will propagate with the velocity of light. We want to have a reasonable mechanism that the fictitious detectors will get the correct relativistic energy $m/\sqrt{1-v^2}$ through the information of the vacuum excitations. At a later time $t=r$,  when the vacuum excitations arrive at the location $\bf r$ of a detector  whose solid angle is $d\Omega$, there are two effects we must consider: the redshift $z=\lambda_{absorption}/\lambda_{emission}-1$ and the decreasing of the rate of arrival of the vacuum excitation waves to the detector. These two effects are similar to the calculations of the apparent luminosity $l$ of a source of absolute luminosity $L$ at a distance $d$ \cite{WeinbBook}. In addition, the intensity of the vacuum excitations should be proportional to $m/\sqrt{1-v^2}$. The effective energy experienced or deduced by this detector is then written as 
\begin{equation}
d\epsilon_{eff}=\frac{m}{\sqrt{1-v^2}}\frac{1}{4\pi r^2(1+z)^2}r^2 d\Omega, 
\end{equation}
with $1+z=(1-v\cos\theta)/\sqrt{1-v^2}$ based on the relativistic Doppler redshift. Here $\theta$ is the angle between $\bf r$ and $\bf v$. When all the detectors on the spherical surface are considered, the overall effective energy is $\epsilon_{eff}=\int d\epsilon_{eff}$. Simple calculations show that $\epsilon_{eff}=m/\sqrt{1-v^2}$ which means that our considerations of the effective energy experienced by the detector inside the spacetime are reasonable. It is also interesting to notice that once the detector gets the information of $\theta$ and $z$, we can deduce the relative velocity of the particle.

To calculate the effective energy of matter for our universe, of course we must consider the evolution of the universe. In this work, we consider the line element for the flat Robertson-Walker metric
\begin{equation}
ds^2=-dt^2+a^2(t)\left[dr^2+r^2(d\theta^2+\sin^2\theta d\phi^2)\right],
\label{ds}
\end{equation}
where $a(t)$ is a function of the time coordinate $t$ called the scale factor. $a_0=a(t_0)$ is the scale factor at the present time and $a_0=1$ with the usual convention. Here $r$ is the  radial distance of the comoving coordinate system.

We assume that the present energy density of vacuum, matter and radiation are 
$\rho_{\Lambda 0}=\Omega_\Lambda\rho_{crit}$,  $\rho_{M0}=\Omega_M\rho_{crit}$, and $\rho_{R0}=\Omega_R\rho_{crit}$, respectively. $\rho_{crit}=3H_0^2/8\pi G$ is the critical present density with $H_0=\dot a_0/a_0$ the Hubble's constant. $\Omega_\Lambda$, $\Omega_M$ and $\Omega_R$ satisfy the condition $\Omega_\Lambda+\Omega_M+\Omega_R=1$, and $\Lambda=3H_0^2\Omega_\Lambda$. Based on Einstein's field equation, we have $\rho_\Lambda(t)=\rho_{\Lambda 0}$, $\rho_M(t)=\rho_{M0}(a_0/a(t))^3$ and $\rho_R(t)=\rho_{R0}(a_0/a(t))^4$. The fundamental Friedmann equation to determine the evolution of $a(t)$ is
\begin{equation}
\left(\frac{\dot a}{a}\right)^2=\frac{8\pi G}{3}(\rho_\Lambda+\rho_M+\rho_R).
\end{equation}

For the detectors inside the spacetime at the present time, similarly to the considerations of the Minkowski spacetime, the detectors will deduce the effective energy of the matter through the vacuum excitations induced by the matter. When the detector measures the vacuum excitations, we should also consider the redshift and the rate of arrival of these vacuum excitations. Because the detection of these vacuum excitations is a local interaction process, the detector can not distinguish whether the redshift and the change of the rate of arrival are due to the expansion of the universe or due to the peculiar motion of the matter.
Hence, when the peculiar motion is negligible and there is a redshift $z$ of the vacuum excitations for the case of expanding universe, it is equivalent that the particle has relativistic energy of $mf_M(z)$ with $f_M(z)=1/\sqrt{1-v^2(z)}$ and $v(z)=((1+z)^2-1)/((1+z)^2+1)$ in a fictitious static universe. 

Based on the above considerations, it is straightforward to find that the overall effective energy experienced by the spacetime at the present time is
\begin{equation}
E_M^{eff}=\int_0^{r_{horiz}}\frac{\rho_{M0}f_M(z)}{(1+z)^2}r^2drd\Omega.
\label{effM}
\end{equation}
Here $r_{horiz}$ is the horizon radius for a detector inside the spacetime at the present time. $d\Omega$ is the differential solid angle. The inclusion of the term $f_M(z)$ in the above equation can be explained further with Einstein's equivalence principle. For the detector comoving with the spacetime, it can not experience gravity and the expansion of the universe. However, the detector will notice the effects of redshift and the change of the rate of arrival of the vacuum excitation waves induced by a particle. The equivalence principle means that the detector will think that the particle has an equivalent relative velocity $v(z)$, which is relevant directly to the redshift $z$.
 
In calculating Eq. (\ref{effM}), we need the relation between $r$ and $z$ to carry out the integral on $r$. From $ds=0$ for the propagation of the vacuum excitations and the fundamental Friedmann equation, we have
\begin{equation}
r(z)=\frac{1}{a_0H_0}\int_{1/(1+z)}^1\frac{dx}{x^2\sqrt{\Omega_\Lambda+\Omega_M x^{-3}+\Omega_R x^{-4}}}.
\end{equation}
$r(z)$ is the radial coordinate of a source that is observed now with redshift $z$.
$r_{horiz}$ is obtained by setting $z\rightarrow\infty$.

\textit{(ii) The effective overall energy of radiation} 

It is similar to calculate the effective overall energy of radiation. We stress that similarly to matter, when the effective energy is addressed, what the detector inside the spacetime experiences is not radiation itself, but through the vacuum excitations induced by radiation. The propagation of the vacuum excitations makes the detector inside the spacetime have the chance of deducing the effective energy. The formula to calculate the effective radiation energy is
\begin{equation}
E_R^{eff}=\int_0^{r_{horiz}}\frac{\rho_{R0}f_R(z)}{(1+z)^2}\times r^2drd\Omega.
\label{radiation}
\end{equation}
Here $f_R(z)=1+z$, which plays a similar role of $f_M(z)$ in the calculations of the effective overall energy of matter. For $z>>1$, the integral factor in the calculations of the effective energy of matter has the property $\rho_{M0}f_M(z)/(1+z)^2\approx \rho_{M0}/2z$, while for the radiation case we have $\rho_{R0}f_R(z)/(1+z)^2\approx\rho_{R0}/z$. This shows the self-consistency of our method to calculate the effective energy for matter and radiation.

There is another way to understand Eq. (\ref{radiation}) to calculate the effective overall energy of radiation. We can calculate the effective radiation energy as follows:
\begin{equation}
E_R^{eff}=\int_0^{r_{horiz}}\frac{\rho_{R}(r(z))}{(1+z)^2}a^3(r(z))\times r^2drd\Omega.
\end{equation}
$\rho_{R}(r(z))=\rho_{R0}/a^4(r(z))$ is the radiation energy density with redshift of $z$ and $a(r(z))=1/(1+z)$. We see that the above equation is the same as Eq. (\ref{radiation}).

\textit{(iii) The effective overall energy of quantum vacuum} 

The stress-energy tensor of the cosmological constant term in a local inertial frame is
\begin{equation}
T^{\Lambda}_{\mu\nu}=-\frac{\Lambda}{8\pi G}\eta_{\mu\nu}.
\end{equation}
Here $\eta_{\mu\nu}$ is the metric of the Minkowski spacetime. Considering another local inertial frame with relative velocity $\bf{v}$, it is easy to verify that $\left(T^{\Lambda}_{\mu\nu}\right)'=T_{\mu\nu}^{\Lambda}$, i.e., they observe the same vacuum energy density. This property is quite different from the cases of matter and radiation. Hence, for an observer inside the spacetime, the overall effective vacuum energy is
\begin{equation}
E_{\Lambda}^{eff}=\int_0^{r_{horiz}}\frac{\rho_{\Lambda 0}f_{\Lambda}}{(1+z)^2}\times r^2drd\Omega.
\end{equation}
Here $f_{\Lambda}=1$. It is worthwhile to mention that although the field fluctuations in the quantum vacuum are random, they satisfy the principle of relativity that in a local inertial frame the field fluctuations observed are the same for every unaccelerated observer \cite{Zeldovich1}. Of course, this strongly suggests the quantum-field origin of the cosmological constant.                                                     

When the quasi-thermal equilibrium between the quantum vacuum and the material contents is considered, based on Postulate 2, the relation between $\rho_{\Lambda 0}$, $\rho_{M0}$ and $\rho_{R0}$ is then determined by
\begin{equation}
E_{\Lambda}^{eff}=E_{M}^{eff}+E_{R}^{eff}.
\label{energycondition}
\end{equation}

\section{Numerical result of $\Omega_\Lambda$}

Usually, the astronomical observations give the parameters $\Omega_\Lambda$, $\Omega_M$ and $\Omega_R$. The best current astronomical observations indicate that $\Omega_\Lambda=68.5\%$ and $\Omega_M=31.5\%$ \cite{data}. The radiation components such as neutrinos and photons contribute a very small amount. With the observation of cosmic background radiation and model of neutrinos, we have $\Omega_R=4.15\times 10^{-5}h^{-2}$ with $h\approx 0.678$\cite{WeinbBook}. 

Assume a dimensionless parameter $r_0=a_0H_0\sqrt{\Omega_M}r$, we have
\begin{equation}
r_0(z)=\int_{1/(1+z)}^1\frac{dx}{x^2\sqrt{\alpha+x^{-3}+\beta x^{-4}}}.
\end{equation}
Here $\alpha=\Omega_\Lambda/\Omega_M$ and $\beta=\Omega_R/\Omega_M$.

From Eq. (\ref{energycondition}), the equation to calculate $\alpha$ is

\begin{equation}
\int_0^{r_{0h}}\frac{f_M(z)}{(1+z)^2}r_0^2dr_0+\int_0^{r_{0h}}\frac{\beta f_R(z)}{(1+z)^2}r_0^2dr_0=\int_0^{r_{0h}}\frac{\alpha f_{\Lambda}(z)}{(1+z)^2}r_0^2dr_0.
\end{equation}
Here
\begin{equation}
r_{0h}=\int_0^1\frac{dx}{x^2\sqrt{\alpha+x^{-3}+\beta x^{-4}}}.
\end{equation}

When $\beta$ is known in advance, we can get $\alpha$ from the above two equations. With the condition $\Omega_\Lambda+\Omega_M+\Omega_R=1$, we can get the dark energy proportion $\Omega_\Lambda$. Radiation becomes important only for high redshifts of $z\geq 10^3$. Hence, it is a good approximation to neglect the contribution of radiation when we calculate $\Omega_\Lambda$. With simple numerical calculations, we get $\Omega_\Lambda\approx 68.85\%$, which is in excellent agreement with $\Omega_\Lambda=68.5\%$ obtained from astronomical observations. It is amazing to notice that we do not adopt any adjustment parameter to get this excellent agreement.

\section{Summary and Discussion}

In summary, based on the considerations of the quantum state of the universe including the vacuum and the assumption of the quasi-thermal equilibrium condition, we calculate in a quantitative way the cosmological constant term and find excellent agreement with the present astronomical observations. Because there is no fitting parameter in our calculations, we have the chance to test further our model with planned 100-fold improvement \cite{SA} in the precision of the measured properties of dark energy in the coming decade. In principle, our theory could calculate the cosmological constant by including the inflationary stage of the universe, and thus provide the chance to get the information of the inflationary stage based on future astronomical observations. Even future works invalidate some of our physical pictures, because the present work provides an amazing coincidence between theory and astronomical observations, we believe that our formula to calculate the cosmological constant will give important clue for future studies, just as Balmer's formula preceded Bohr's theory and Planck black-body radiation law preceded energy quantization. 

Of course, the present work only provides a phenomenological theory to calculate the cosmological constant. It is not the purpose of this work to consider the final solution of the unification of gravity and quantum mechanics. Nevertheless, our theory gives a new idea to solve the fine tuning problem and coincidence problem for the dark energy density by noting that when the radiation is omitted, the dark energy proportion in our theory is a universal value for flat universe. At difference cosmic time, although the dark energy density observed by a real observer can be different, the dark energy proportion is still the same. This is the reason why we firstly calculate the dark energy proportion and then we can get the cosmological constant. It is natural that people will propose another type of the coincidence problem that why our phenomenological theory agrees excellently with the astronomical observations. It is not clear for this question at the present stage and this question may provide the starting point for future studies.

At last, we further discuss a cosmic evolution model of an initial quantum universe without matter and radiation. Without matter and radiation, both the average vacuum energy and energy fluctuation at the sale of the Planck length are of the order of the Planck energy. For this initial state of the universe, we may envisage two stages of the universe evolution illustrated in Fig. 1. (1) Inflationary stage with the creation of matter and radiation. In this inflationary stage, the scale factor expands exponentially rapidly, and the vacuum energy fluctuations are so large that matter and radiation are created from the quantum vacuum. During this inflationary stage, however, there is no stable thermal equilibrium between the quantum vacuum and the material contents of the universe, and the quantum vacuum can not be approximated as the superposition of the states of different cosmological constants. Gradually, the inflationary stage ends because of the establishment of the quasi-thermal equilibrium between the quantum vacuum and the material contents. In this case, the fluctuation of the vacuum energy becomes small, while the vacuum energy density is of the order of the material contents. After the fluctuation of the vacuum energy is highly suppressed, the creation of matter and radiation from the quantum vacuum will become negligible. The gradual creation of the material contents and the gradual establishment of the quasi-thermal equilibrium lead to the second stage of our universe, i.e.,  (2) Quasi-thermal equilibrium stage between the quantum vacuum and the material contents. In this stage, the quantum vacuum becomes the superposition of states of different cosmological constants. The cosmological constant with the maximum probability is determined by the quasi-thermal equilibrium condition. The cosmic history observed by a real observer is then determined by this cosmological constant based on Everett's relative-state formulation. 

\section*{Acknowledgements}
We thank the numerical calculations by Xuguang Yue and Yunuo Xiong. This work was supported by NSFC 11175246, 11334001.


\begin{thebibliography}{10}
\expandafter\ifx\csname url\endcsname\relax
  \def\url#1{\texttt{#1}}\fi
\expandafter\ifx\csname urlprefix\endcsname\relax\def\urlprefix{URL }\fi
\providecommand{\bibinfo}[2]{#2}
\providecommand{\eprint}[2][]{\url{#2}}

\bibitem{SA}
\bibinfo{author}{A. G. Riess}, and \bibinfo{author}{L. Mario},
\newblock \bibinfo{title}{\textit{The Puzzle of Dark Energy}},
\newblock \bibinfo{journal}{Sci. Am.}
  \textbf{\bibinfo{volume}{314}}, \bibinfo{pages}{38} (\bibinfo{year}{2016}).

\bibitem{acceleration}
\bibinfo{author}{A. G. Riess}, et al.,
\newblock \bibinfo{title}{\textit{Observational Evidence from Supernovae for an Accelerating Universe and a Cosmological Constant}},
\newblock \bibinfo{journal}{Astron. J.}
  \textbf{\bibinfo{volume}{116}}, \bibinfo{pages}{1009} (\bibinfo{year}{1998}).

\bibitem{acceleration1}
\bibinfo{author}{S. Perlmutter}, et al.,
\newblock \bibinfo{title}{\textit{Measurements of Omega and Lambda from 42 High-redshift Supernovae}},
\newblock \bibinfo{journal}{Astron. J.}
  \textbf{\bibinfo{volume}{517}}, \bibinfo{pages}{565} (\bibinfo{year}{1999}).

\bibitem{weinberg}
\bibinfo{author}{S. Weinberg},
\newblock \bibinfo{title}{\textit{The Cosmological Constant Problem}},
\newblock \bibinfo{journal}{Rev. Mod. Phys.}
  \textbf{\bibinfo{volume}{61}}, \bibinfo{pages}{1} (\bibinfo{year}{1989}).


\bibitem{cos}
\bibinfo{author}{P. J. E. Peebles}, and \bibinfo{author}{B. Ratra},
\newblock \bibinfo{title}{\textit{The Cosmological Constant and Dark Energy}},
\newblock \bibinfo{journal}{Rev. Mod. Phys.}
  \textbf{\bibinfo{volume}{75}}, \bibinfo{pages}{559} (\bibinfo{year}{2003}).


\bibitem{quin}
\bibinfo{author}{R. R. Caldwell}, and \bibinfo{author}{R. Dave}, and \bibinfo{author}{P. J. Steinhardt},
\newblock \bibinfo{title}{\textit{Cosmological Imprint of an Energy Component with General Equation of State}},
\newblock \bibinfo{journal}{Phys. Rev. Lett.}
  \textbf{\bibinfo{volume}{80}}, \bibinfo{pages}{1582} (\bibinfo{year}{1998}).

\bibitem{Zeldovich}
\bibinfo{author}{Y. B. Zel'dovich},
\newblock \bibinfo{title}{\textit{Cosmological Constant and Elementary Particles}},
\newblock \bibinfo{journal}{JETP. Lett.}
  \textbf{\bibinfo{volume}{6}}, \bibinfo{pages}{316} (\bibinfo{year}{1967}).


\bibitem{Zeldovich1}
\bibinfo{author}{Y. B. Zel'dovich},
\newblock \bibinfo{title}{\textit{Cosmological Constant and Theory of Elementary Particles}},
\newblock \bibinfo{journal}{Sov. Phys. Usp.}
  \textbf{\bibinfo{volume}{11}}, \bibinfo{pages}{381} (\bibinfo{year}{1968}).


\bibitem{Wheeler}
\bibinfo{author}{J. A. Wheeler},
\newblock \bibinfo{title}{\textit{On the Nature of Quantum Geometrodynamics}},
\newblock \bibinfo{journal}{Ann. Phys.}
  \textbf{\bibinfo{volume}{2}}, \bibinfo{pages}{604} (\bibinfo{year}{1957}).

\bibitem{multiverse}
\bibinfo{author}{S. Kachru}, \bibinfo{author}{R. Kallosh}, \bibinfo{author}{A. Linde}, and \bibinfo{author}{S. P. Trivedi},
\newblock \bibinfo{title}{\textit{De Sitter Vacua in String Theory}},
\newblock \bibinfo{journal}{Phys. Rev. D}
  \textbf{\bibinfo{volume}{68}}, \bibinfo{pages}{046005} (\bibinfo{year}{2003}).


\bibitem{Susskind}
\bibinfo{author}{L. Susskind},
\newblock \bibinfo{title}{\textit{The Anthropic Landscape of String Theory}},
\newblock \bibinfo{journal}{arXiv:hep-th}
\bibinfo{pages}{0302219} (\bibinfo{year}{2003}).

\bibitem{Everett}
\bibinfo{author}{H. Everett},
\newblock \bibinfo{title}{\textit{Relative State Formulation of Quantum Mechanics}},
\newblock \bibinfo{journal}{Rev. Mod. Phys.}
  \textbf{\bibinfo{volume}{29}}, \bibinfo{pages}{454} (\bibinfo{year}{1957}).


\bibitem{Dewitt}
\bibinfo{author}{B. S. Dewitt},
\newblock \bibinfo{title}{\textit{Quantum Theory of Gravity. I. The Canonical Theory}},
\newblock \bibinfo{journal}{Phys. Rev.}
  \textbf{\bibinfo{volume}{160}}, \bibinfo{pages}{1113} (\bibinfo{year}{1967}).


\bibitem{Misner}
\bibinfo{author}{C. W. Misner},
\newblock \bibinfo{title}{\textit{Quantum Cosmology. I}},
\newblock \bibinfo{journal}{Phys. Rev.}
  \textbf{\bibinfo{volume}{186}}, \bibinfo{pages}{1319} (\bibinfo{year}{1969}).


\bibitem{Hawking}
\bibinfo{author}{J. B. Hartle}, and \bibinfo{author}{S. W. Hawking},
\newblock \bibinfo{title}{\textit{Wavefunction of the Universe}},
\newblock \bibinfo{journal}{Phys. Rev. D}
  \textbf{\bibinfo{volume}{28}}, \bibinfo{pages}{2960} (\bibinfo{year}{1983}).



\bibitem{MyWork}
\bibinfo{author}{H. W. Xiong},
\newblock \bibinfo{title}{\textit{Repulsive Gravitational Effect of a Quantum Wave Packet and Experimental Scheme with Superfluid Helium}},
\newblock \bibinfo{journal}{Front. Phys.}
  \textbf{\bibinfo{volume}{10}}, \bibinfo{pages}{100401} (\bibinfo{year}{2015}).

\bibitem{WeinbBook}
\bibinfo{author}{S. Weinberg},
\newblock \bibinfo{title}{\textit{Cosmology}},
\newblock \bibinfo{journal}{Oxford (New York)}
 (\bibinfo{year}{2008}).

\bibitem{data}
\bibinfo{author}{P. A. R. Ade}, et al.,
\newblock \bibinfo{title}{\textit{Planck 2015 Results XIII. Cosmological Parameters}},
\newblock \bibinfo{journal}{Astron. Astrophy.}
  \textbf{\bibinfo{volume}{594}}, \bibinfo{pages}{A13} (\bibinfo{year}{2016}).



\end{thebibliography}
\end{document}